# Improving the Reliability of Next Generation SSDs using WOM-v Codes

Shehbaz Jaffer[1,2], Kaveh Mahdaviani[1], and Bianca Schroeder[1]

[1]*University of Toronto*, [2]*Google*

## Abstract

*High density Solid State Drives, such as QLC drives, offer increased storage capacity, but a magnitude lower Program and Erase (P/E) cycles, limiting their endurance and hence usability. We present the design and implementation of non-binary, Voltage-Based Write-Once-Memory (WOM-v) Codes to improve the lifetime of QLC drives. First, we develop a FEMU based simulator test-bed to evaluate the gains of WOM-v codes on real world workloads. Second, we propose and implement two optimizations, an efficient garbage collection mechanism and an encoding optimization to drastically improve WOM-v code endurance without compromising performance. A careful evaluation, including microbenchmarks and trace-driven evaluation, demonstrates that WOM-v codes can reduce Erase cycles for QLC drives by 4.4x-11.1x for real world workloads with minimal performance overheads resulting in improved QLC SSD lifetime.*

## 1 Introduction

Flash-based Solid State Drives (SSDs) offer a faster alternative to Hard Disk drives (HDDs), but have a major limitation: unlike HDDs, where previously written data is over-writable, a flash cell needs to be erased before it can be programmed, and each erase operation causes wear-out that reduces a cell's lifetime. Older generations of flash were based on single-level cells (SLC), which store only a single-bit in a cell and can typically tolerate several thousand program and erase cycles before wearing out. However, to keep up with the increasing demand for storage capacity, more bits need to be stored in a cell. Such SSDs are called multi-bit cell SSDs. Recent work [26] shows that with each additional bit stored in one SSD cell, the number of erase cycles that the SSD can endure reduces by an order of magnitude. Figure 1 illustrates the problem based on recent projections [5]. Flash based on Multilevel Cells (MLC) and Triple Level Cells (TLC), which are common nowadays, can tolerate a significantly smaller number of P/E cycles. Recently, QLC drives have started being deployed in datacenters. Even more worrisome is a look into the future with PLC drives, which might see P/E cycle limits drop to tens or a few hundred. To make high-density SSD drives usable beyond archival applications, it is paramount to reduce the number of times the storage media is erased.

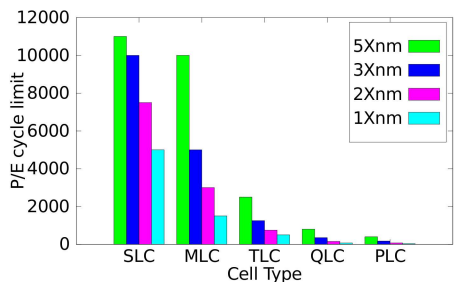

Fig 1:*Reduced P/E cycles with increased storage density [5]*

In our recently published workshop paper [12] we have shown great promise in improving endurance of multi-bit cell SSDs using WOM-v codes. WOM-v codes use a single lookup table to provide a low-performance overhead encoding scheme during a write and a read operation. While there exist other codes that allow additional overwrites before erase and are more space optimal, such codes traditionally involve multiple iterations before encoding and decoding is completed and are therefore performance inefficient. Further, WOM-v codes provide a family of codes that can be adapted to the amount of space overheads the underlying storage media can tolerate. In [12], we present theoretical back-of-the-envelope calculations based on a simplistic cell level model. Our calculations show that upto 500% additional writes can be done on QLC drives using WOM-v codes [12].

Although the theoretical results show significant improvement, it is not clear whether these improvements can be achieved in practice. First, in the real world, SSDs have several restrictions while overwriting data. SSD contains multiple erase blocks containing multiple pages. Writes to erase blocks is done at page granularity, where pages can be written to in sequential order from the first to the last page. Only a limited number of erase blocks can be programmed at a time. Unless all the pages within the target erase block are invalidated, the SSD must first relocate all valid pages from target erase block to another erase block. Once the valid pages are relocated, the entire erase block is available to be reprogrammed. This "copy before overwrite" step, also called garbage collection, considerably increases the overall writes done to the device, which counters the gains we get from WOM-v codes. The evaluation in [12] falls short as it ignores the garbage collection workflow in SSDs.

Second, modern SSDs employ parallelism for higher performance. Multiple erase blocks are arranged in groups called *Erase Units (EUs)*. Only one EU is active at a time. Incoming data is first buffered and subsequently sharded across all erase blocks in the active EU. Any performance based metric computation may not capture such nuances involved in writing data to a shared buffer that requires preventing race conditions using locks, nor the gains due to striping the data across different parallel units. The evaluation in [12] fails to do any performance evaluation, and hence needs a real world flash emulator to evaluate the impact of WOM-v code on application performance.

Finally, real world workloads vary in their storage access patterns. The access pattern determines the amount of garbage collection in the device. A thorough analysis of WOM-v codes over multiple real world workloads is required to assess the practical gains of WOM-v codes introduced by [12].

The contribution of our paper is to demonstrate that WOM-v codes have real gains in practice. First, we present the first paper to our knowledge that provides a detailed design, implementation and evaluation of Non-Binary WOM codes on next generation dense SSD (QLC) drives to improve SSD lifetimes. Second, we show that there is a difference in theory and practice in the amount of SSD Erase Cycle reduction that can be achieved using WOM-v codes. With a system implementation and evaluation, we see more realistic gains than those provided by a theoretical evaluation. Moreover, counter intuitive to theoretical assumptions, we show that higher order WOM-v codes do not provide Erase Cycle reduction if the workload generates high write amplification. For high write amplification workloads, we provide two novel optimizations- GC-OPT and NR-Mode - which significantly reduce erase operations while retaining high performance. Third, our simulator is open-sourced and can be used as a test-bed to evaluate future WOM code designs on next generation SSDs. Finally, we show how WOM-v codes improve the lifetime of QLC flash by 4.4x-11.1x with negligible performance overheads.

## 2 Background

Traditional WOM codes were first proposed in the 1980s for media such as punch cards, where data is written bit-wise and each written bit can only be changed in one direction, e.g. from 0 to 1 [24]. More recently, WOM codes have attracted attention since their model of changing a written bit in only one direction matches the characteristics of a (single-level) flash cell. Prior work uses this observation by applying WOM codes to increase the lifetime of flash [32]. We refer to these WOM codes, which assume bits can only be changed from 0 to 1 as *binary-WOM* codes.

In our workshop paper [12] we observe that these binary WOM codes are a poor fit for new generations of flash cells, such as MLC or QLC, where more than one bit is encoded in a single cell. The real constraint for a flash cell is that the voltage level of a cell can always be increased (up to some maximum level $V_{max}$), but not decreased, independently of what the encoded bit values are. Adhering to a binary model creates unnecessary constraints that limit the power of a codes and also does not match device characteristics.

To quantify the limitations of a binary WOM codes for a QLC drive, we write a search program to compute the maximum number of generations writable using binary WOM(2,4) encoding scheme. We find that we are unable to write more than 2 generations of data using WOM(2,4) coding scheme. Hence, beyond two overwrites on a QLC drive, an erase operation will be required. Moreover, the code will have a 2x space amplification since 2 bits of data are encoded into 4 bits of codeword. Hence there is no net gain in using Binary WOM codes for QLC drives. Furthermore, if Binary WOM codes are implemented in a real SSD, the additional writes due to garbage collection will perform worse than not using WOM code at all. Hence in the remaining paper, we compare non-binary WOM-v(k,N) codes with non-encoded (NO-WOM) configuration.

Instead [12] proposes a new family of WOM codes for QLC flash, referred to as *voltage-based* WOM codes (WOM-v codes), that are based on the voltage level constraint and achieve a higher number of overwrites between erases.

For a detailed description of WOM-v codes for QLC flash, we refer the reader to [12]. However, for convenience we provide a high-level summary of how these codes work in the remainder of this section.

### 2.1 Introduction to WOM-v Codes

A WOM-v code has two parameters *x* and *y* and a WOM-v(x,y) code encodes x bits of data into a code word of y bits. In the case of QLC flash, y is equal to 4 and the $2^4$ code words correspond to the 16 voltage values a QLC flash cell can have. Figure 2 shows an illustration of the three specific WOM-v codes for QLC flash [12] presents: WOM-v(3,4), WOM-v(2, 4) and WOM-v(1,4).

For a simplified explanation of WOM-v codes consider the WOM-v(3,4) code on the left in Figure 2. The column labelled "CODE" shows the mapping of code words to a cell's voltage levels V0 to V15, where V0 is the lowest and V15 is the highest voltage level. The left column labelled "DATA" shows how the 3-bit data words are mapped to the 16 voltage levels.

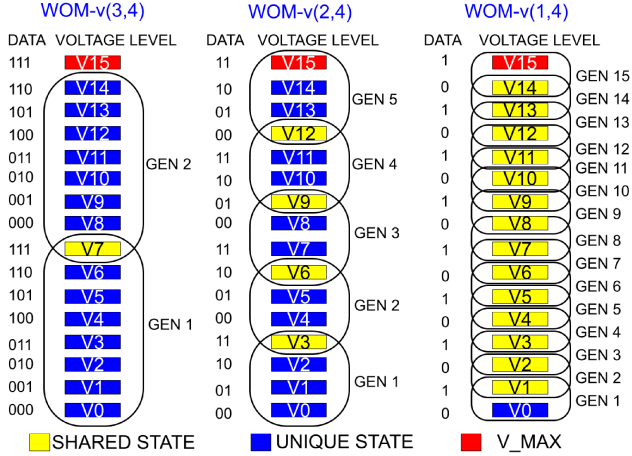

Figure 2: *Voltage based codes for encoding 3 2 and 1 data bit(s) into 16 Voltage Levels. Each oval represents a generation. Each generation contains a set of Voltage Levels mapped to a set of all possible data bit sequences for a coding scheme.*

Note that each 3-bit data word appears twice in the "DATA" column, once mapped to the bottom half of voltage levels and once mapped to the top half of voltage levels. We refer to the bottom half of voltage levels as generation 1 (GEN1) and the top half as generation 2 (GEN2). Having for each data word a corresponding mapping in each generation ensures that we can write to a freshly erased cell as many times as we have generations without erasing. To see how consider again the example of the WOM-v(3,4) code: If we ensure that the first write to a cell (after an erase) maps the data word to the bottom half of the voltage levels (GEN1), then we can map the second write of any data word to the top half of voltage levels (GEN2). Note that writing to a higher generation only requires increasing a cells voltage, which can be done without an erase. For example, sequentially writing the two data words 101 and 001 to a freshly erased cell would involve programming the cell first to the voltage level corresponding to V5 and V9 respectively. We define each iteration of write to the underlying device as one *write cycle*.

## 2.2 Optimizations in WOM-v Codes

While the above describes the basic idea behind WOM-v codes, our workshop paper [12], describes 3 optimizations that further improve the efficiency of WOM-v codes:

**1. Same-generation transitions:** Sometimes it will be possible to write a data word without increasing voltage levels into the next generation. Consider the example of sequentially writing the two data words 001 and 101 to a freshly erased WOM-v(3,4) cell. Writing 001 raises the voltage to the level corresponding to V1 in GEN1. Now observe that the second data word, 101, has a mapping within GEN 1 (V5) that corresponds to a higher voltage level than the current voltage level (V1). In this case we do not need to transition to GEN2, but can perform the second write while staying at a GEN1 voltage level. In this case we will be able to write to the cell more often than the number of generations.

**2. Code word sharing:** The reader might have noticed that voltage level V7 in WOM-v(3,4) cell is shared between GEN1 and GEN2. The data word 111 maps to V7 in both GEN1 and GEN2. This sharing allows us to squeeze in more generations. The savings become more obvious when considering WOM-v(2,4), where sharing allows us to create 5 generations instead of otherwise only 4.

**3. Using ECC to increase page writes between erases:** As writes happen at the granularity of pages (not individual cells) we can no longer overwrite a page once any one of its cell has reached its maximum generation. To continue writing to a page with a few cells in the max generation without erasing [12] suggests to make use of pre-existing device error correcting code (ECC). Since flash media are naturally prone to bit errors, in particular with increasing age, any flash-based SSD incorporates ECC. The idea is that if there is only a small number of cells in a page that have reached the max generation and therefore cannot be written to and we had a way to mark these cells as *invalid* while writing to the rest of the page, a later read could rely on the SSD's existing ECC to determine the value of these cells.

An obvious question is whether using existing ECC to reconstruct the content of cells that are marked *invalid* (because they had reached the max generation and could not be written to) will affect reliability or performance of the drive. We make the case that if done right neither reliability nor performance will be affected. The first key observation is that bit error rates of SSDs grow with age/wear-out and that the ECC is provisioned to be able to handle the high bit error rates toward the end of the drive's lifetime. That means that for the majority of a drive's lifetime the ECC is over-provisioned - it is stronger than what is required to ensure drive reliability. Therefore during the first years of a drive's operation, when bit error rates are lower than what the ECC is designed for, the ECC can handle the correction of a certain number of invalid cells without any impact on reliability. To control the impact on reliability we set a threshold called *ECC_threshold* on the fraction of cells per page that are allowed to be marked as invalid before an erase operation is required. If the number of cells in a page that are in their maximum generation reaches this ECC_threshold, further overwrites are not allowed and the page needs to be erased first. In practice, this threshold could be chosen dynamically based on age or observed bit error rates of the SSD.

The second key observation with respect to reliability is that recovering the value of invalid cells is easier than recovering from random bit errors that the ECC is designed to handle. Since the location of the invalid cells is known (in contrast to the unknown random location of bit errors) the ECC only needs to perform erasure *correction* rather than detection and correction. Since correcting *x* erased bits requires half the redundancy as detecting and correcting $x$ erroneous bits [10], it is less likely for the ECC to fail to reconstruct invalid cells.

The discussion above should also make it clear that our use of the existing ECC should not or only minimally affect read latency or throughput. The reduction in performance due to correcting reads of $V_{max}$ cells is the same as that of correcting a corrupted bit. Finally, the use of ECC to enhance WOM-v code performance is optional, WOM-v code can work with lower endurance and maintain similar performance as No-WOM configuration if the underlying device ECC is not used.

### 2.3 Flash Friendliness of WOM-v(k,N) codes

Besides reducing the number of required erase operations, WOM-v codes also have the added benefit that their writes are more flash friendly as they only involve voltage level increases within a short range of voltage levels. For instance, in a standard QLC drive with 16 distinct voltage levels, the voltage increment for a NO-WOM configuration could be anywhere between V0 to V15. However using a WOM-v scheme, the voltage will only monotonically increment by a factor of one generation (eg. the next 4 voltage levels in WOM-v(2,4) coding scheme). In general, for a WOM-v(k,N) coding scheme, the next write will only increase the voltage level from 0 to the next $2^k - 1$ levels as compared to No-WOM configuration where the voltage may increase anywhere between 0 to $2^N$ levels. The implications of this lower rate of increment in voltage level during a write operation for WOM-v(k,N) code is that less amount of charge will be injected to each cell as compared to NO-WOM configuration. This mode of programming a cell in WOM-v encoding scheme is more flash friendly as it induces less program disturb errors in both programmed cell and neighbouring cells as compared to NO-WOM configuration. Further, gradual voltage increment may also simplify SSD circuitry as the number of possible transitions on each state is significantly reduced in WOM-v configuration as compared to the NO-WOM configuration.

In this work, we focus our attention on reducing the number of erase operations in exchange for more number of writes to improve SSD endurance. We acknowledge that there are other factors, including write operations, and temperature that would impact the endurance of SSD drives. However, such factors have much less impact on the endurance of SSD drives as compared to the erase operation [11].

## 3 System Implementation

This section presents the system implementation that we use to evaluate the real world potential of WOM-v codes. With a systems implementation of the WOM-v code, we are able to measure the practical reduction in the number of Erase Units (EUs) erased in SSDs. We are also able to measure the impact of input data contents, workload patterns and performance overheads of WOM-v codes that cannot be accurately measured in theory. This is because SSDs are programmed at the granularity of pages and not cells. Further, SSDs are erased at the granularity of EUs before additional data can be written on them. Ideally, we would want to implement WOM-v code on a real world hardware by manipulating the device Flash Translation Layer(FTL). However, we have the following challenges: 1) SSD FTL is a proprietary closed sourced software that is not available for change. 2) An evaluation done on one hardware device configuration may not be conclusive and applicable to future SSD generations.

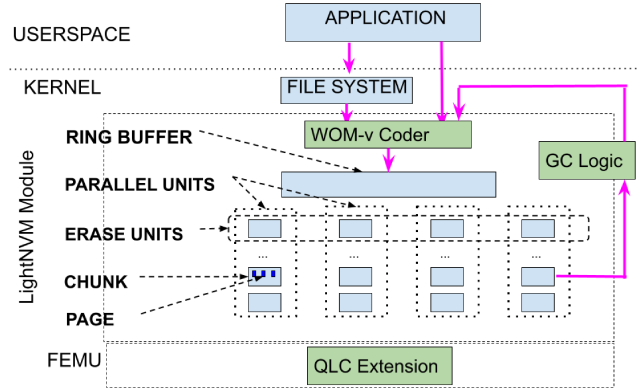

Figure 3: *LightNVM Architecture with our contributions in green. The LightNVM and FEMU modules together emulate a host-managed-SSD device. Our WOM-v encoder logic is inside the LightNVM module. We change the garbage collection workflow to perform erase operations based on WOM-v configuration type. We add QLC-flash support to FEMU.*

To address these issues, we implement WOM-v code in the Linux LightNVM Open-Channel SSD Subsystem module [9], which allows making changes in the device FTL. We add `445` LOC in the LightNVM module and `220` LOC in FEMU. In order to emulate a QLC device, we extend FEMU, a widely used Flash Emulator [17] for MLC devices, to emulate a QLC device. Our extension for QLC support is already merged into the mainline FEMU repository [6]. The WOM-v implementation requires no changes in the application or file system running on top of the device. The computation overheads are negligible as encoding and decoding involve simple table lookups. Although the WOM-v coding scheme is built within the LightNVM module, in reality, some of the key functionality might be implemented using special purpose hardware. We leave more efficient hardware designs of encode and decode operations as part of future work.

### 3.1 LightNVM Architecture

LightNVM is a Linux module that exposes the underlying architecture of a real or an emulated NVMe SSD to the host. This helps us to make modifications in the way we write to and read from the device. LightNVM also enables us to control the garbage collection scheme and when an erase operation should be performed on the underlying device. The internal architecture of LightNVM is as shown in Figure 3. The two main data structures of a LightNVM module are 1) a shared ring buffer and 2) Parallel Units.

### 3.1.1 Ring Buffer

The ring buffer is a circular buffer where data is placed before being written to the underlying device. The device may be accessed either directly by the application or through the file system as shown in Figure 3. Once the ring buffer is full or the user requests a sync operation, the data copied to the ring buffer is striped across different Parallel Units (see 3.1.2) in a round robin order. The ring buffer is a shared resource between two threads, the *user-write thread* that copies the incoming application/file-system data to the ring-buffer and the *gc thread* that copies valid pages from the underlying device to the ring-buffer during garbage collection. The ring buffer and the device configuration remains unchanged across NO-WOM and WOM-v(k,N) configurations.

### 3.1.2 Parallel Units

Figure 3 shows a device with 4 Parallel Units. A Parallel Unit(PU) is an independent unit of storage on the device. Each Parallel Unit is divided into multiple erase blocks or *chunks*. Each *chunk* contains a linear array of *pages* that can be sequentially programmed from the first page to the last page of the chunk. A group of same-sized chunks, one chunk from each Parallel Unit, forms an *erase unit (EU)*. Pages within a chunk are sequentially programmed. Pages across chunks within an EU are programmed in parallel. As a result, all chunks in an EU get filled at the same time. Furthermore, all chunks in an EU can only be erased together and hence are garbage collected at the same time in the default setup. At any time, a single EU is opened for application writes. The EU is closed once all pages in the erase-unit have been programmed.

We can issue 3 types of operations to each parallel unit - 1) page read 2) page write and 3) chunk erase which is issued in parallel to all chunks within an EU. Reads on the emulated device are 10 times faster than writes, and erase operations are 10 times slower than writes.

All operations are performed sequentially within a Parallel Unit. Two operations on different Parallel Units can be performed in parallel. The number of parallel units on an emulated LightNVM module is configurable. We use the default 4 Parallel Units for all our experiments.

It is important to note that our implementation of WOM-v codes does not make any changes to how parallelism across Parallel Units or sequentiality within parallel units works. In 4.2.3 we show that the performance gains provided by LightNVM parallelism are retained even after using WOM-v(k,N) codes.

### 3.1.3 Write and Read Operation

All page writes are staged on the ring buffer. If the data available in the ring buffer is small and a *sync* command is issued by the user, the data to be written is appropriately padded for alignment and striped across Parallel Units. At any give time a single EU, called the *active EU*, is open for writes. Equal number of pages are simultaneously written to all chunks of an active EU, until the last page of all chunks have been programmed. Once the *active EU* has been filled, the EU is closed and a new EU is made *active* and opened for future writes.

All read operations are sent to the device as a block I/O (*bio*) request. The LightNVM module first translates the logical block address (LBA) of the requested page from the *bio* structure to the device Physical Page Address (PPA) using the Logical-to-Physical (L2P) Map. The page contents are then copied from the device PPA to the *bio* request and returned back to the user.

### 3.1.4 Garbage Collection

LightNVM employs garbage collection to reclaim space occupied by invalidated pages. A page gets either explicitly invalidated by a TRIM command, or implicitly because the logical page stored in it gets overwritten by the application. To free SSD pages occupied by such invalidated pages, a closed EU is opened in the *gc-mode* by the *gc-thread* for garbage collection. In the garbage collection phase, all valid pages from a *gc-erase-unit* are copied to the ring buffer. The entire EU is then erased and closed. This EU is returned back to the free pool of erase-units available for the *user-write* thread to be opened for future writes.

LightNVM follows a greedy approach to select an EU to be garbage collected. An EU with the maximum number of invalidated pages is chosen first. LightNVM reserves an over provisioned space of 11% in order to not run out of space while performing garbage collection. Additionally, LightNVM has no wear-leveling mechanism and delegates wear-leveling to lower level drive FTL.

## 3.2 WOM-v Implementation

To incorporate WOM-v codes in the LightNVM code, first, all writes to the device need to be encoded. Second, all reads issued to the device need to decode previously written data. Third, the default garbage collection logic needs to be modified. Instead of erasing all the EUs during garbage collection, an erase should now be selectively done based on the state of the pages within an EU. Fourth, for our experiments, the underlying device emulator needs to support next generation

SSD devices with QLC or denser flash medium. Finally, we implement two optimizations, which while they do not change the design of WOM codes, help to improve the performance and reduce the overheads of WOM coding. We first present the baseline implementation without these improvements in the next subsection and then present the optimizations in sub-section 3.2.2.

### 3.2.1 Baseline Implementation

We add the following components to the LightNVM module (highlighted in green in Figure 3): 1) encode and decode logic 2) WOM-v aware garbage collection logic and 3) QLC support for FEMU. Our framework is extendable to emulate future SSDs and future coding schemes.

**Write Operation**

An application or a file system can submit a write request to LightNVM. All writes are encoded before being written to the drive. By default, a WOM coding scheme first reads the previously written data on the media. This data is encoded and overwritten on the physical page, maintaining the voltage based constraint of the underlying media. Since this default methodology causes increased read amplification, we present a simple mechanism to avoid such reads altogether for WOM-v codes using the No-Reads configuration as discussed in 3.2.2.

During a write operation, the ring buffer creates a mapping between the logical block address (LBA) of pages staged in the ring buffer to the destination physical page address (PPA) of the pages on the device before writing the pages to the device. We intercept all writes at this stage and apply the following transformation: First, we read preexisting encoded data in the PPA of all pages being written. (In 3.2.2 we describe a No-Reads configuration that obviates the need for this read operation). Next we encode incoming pages using the preexisting data. The encoding scheme is straightforward and involves a simple lookup in the static WOM-v(k,N) encode table shown in Figure 2. Finally, we write the new encoded pages to the device PPA on the drive.

For a WOM-v(1,4) coding scheme, each page incoming write is encoded and stored in 4 x 4KB physical of a QLC page on the device. Similarly, for a WOM-v(2,4) coding scheme, each 4KB incoming write is encoded and stored in 2 x 4KB physical pages on the device. To reduce the performance overheads of additional page writes, we increase the logical page size at which an application page is sent to the device after an encode operation to 16KB and 8KB for the WOM-v(1,4) and WOM-v(2,4) configuration respectively. We maintain logical page locality among all encoded pages. i.e. all pages belonging to the original logical page are encoded into consecutive logical LBAs. We describe the importance of logical page locality during reads in the next sub-section.

**Read Operation**

In the read workflow, the original read block I/O (*bio*) request from the application is first translated into the corresponding consecutive LBA address *bio* requests. The consecutive pages correspond to a single page due to logical page locality. Next, all encoded pages that were read are decoded in the read return path. The decoded data is copied to the originally submitted *bio* request structure and can be read by the application with no modifications. The decoding scheme is straightforward and involves a simple lookup in the static WOM-v(k,N) encode table shown in Figure 2.

**Garbage Collection**

Like the standard LightNVM garbage collection mechanisms, our implementation chooses the erase unit with the smallest number of valid pages for garbage collection and copies out all valid pages to the ring buffer. However, while the standard scheme would now erase the erase unit, our modified scheme will erase the erase unit only if any of its pages (valid or invalid) have reached the *ECC_threshold* on the number of cells in the maximum generation (recall Section 2.2).

We set the *ECC_threshold* to 3% in our experiments, i.e. we erase an erase-unit that has any page with more than 3% cells in *GEN_MAX*. This threshold is chosen based on the theoretical evaluation in [12] combined with the fact that current devices have reported 7% ECC in each page [23]. We predict higher ECC being reserved for QLC and future generation drives.

**Cell and Page Layout**

In a standard SSD based on N-level cells each cell can be programmed to take on one of $2^N$ different voltage levels. Our work assumes the same type of cells as a standard SSD with the same number of voltage levels and mechanisms for performing voltage increments. However, we change how these cells and voltage levels are used to store data. Besides the encoding of data as described in Section 2, we also change how cells are assigned to pages. Conventionally, for a N-level cell drive, 1 cell stores N-bits of data, where each bit is located on N different pages. For example, in QLC flash, 1 cell stores 4 bits of information, and each bit is located in 4 different pages of the drive. The drawback of this approach is that the 4 pages have to be programmed in a certain order. For WOM-v codes, we propose that instead of mapping each cell to 4 bits in 4 different pages, map 4 bits of information representing 1 cell in a single page. This mapping helps us program a single cell of the page to a voltage value of our choice. Using this technique, for a WOM-v(2,4) code, we will be able to encode 1 4KB logical page into 8KB data and store this data in 2K cells of the flash media. Similarly, for a WOM-v(1,4) code, 1 logical page of 4KB will get encoded as 16KB of data and be written to 4K cells of the flash media. We also increase the logical page size to 2x and 4x the size of original logical page size for WOM-v(2,4) and WOM-v(1,4) configurations respectively. The implication of the above approach is that we no longer have to maintain a specific sequence or order of page programming. Instead, each cell stored on a single page can be programmed to a specific voltage range determined by the generation of the cell in the write cycle.

A WOM-v(k,N) scheme can be naturally extended to any N-level cell by changing the value of N to the number of bits each cell of the device. The value of k determines the space-endurance tradeoff - a lower value of k will give higher endurance but also consume more physical space.

### 3.2.2 WOM-v Optimizations

We identify two optimizations to the baseline implementation of WOM-v(k,N) codes. First, we present `GC_OPT Mode`, a novel methodology for garbage collection in WOM-v(k,N) configuration that improves SSD endurance considerably by delaying valid page rewrites during garbage collection. Second, we present `NR Mode` WOM coding scheme, a technique to perform encode operations and overwrite an invalidated page without reading the previously existing contents of the page which completely eliminates read amplification during writes.

#### GC_OPT Mode

The goal of this optimization is to reduce the write-amplification caused by copying out valid pages from an EU during garbage collection. The key observation is that with WOM-v codes in most cases the invalid pages in an EU can be over-written again, without first erasing them. Recall from Section 3.2.1 that we make use of this fact and only perform an erase when a page in an EU has reached its ECC_threshold. That means that in those cases where no erase is required for an EU, we can leave the valid pages in place (without copying them out), *as long as we skip writing to valid pages* when writing to this EU in the future. (The remaining, non-valid pages within a chunk will be overwritten in the same specific order as before in order to minimize cell-to-cell interference, and writes are parallelized across the parallel units of an EU, as before.)

#### No-Read (NR) Mode

The WOM-v codes we introduce share a source of potentially major performance overheads with other WOM codes previously proposed: to write to a cell we need to know the cell's contents in order to encode the data to be written. Therefore each write necessitates a prior read. In this subsection we make two observations about WOM-v(k,N) codes that allow us to eliminate this read-before-write: (1) The only reason we need to read a cell's contents before writing to it, is to determine which generation to use for the write. (2) If we remove the same generation transition optimization (recall Section 2) and instead in each write cycle move to the next generation, we can store the most recently used generation for each page with a chunk's metadata. When opening an EU for writing, this metadata can be loaded into memory and all writes to pages in the EU are done using the generation information in the metadata (obviating the need for reading the page). We refer to the above method for eliminating the read-before-write as *NR (No-Read) mode*.

We note that NR mode has two potential downsides: First, same generation transitions are no longer possible, which might reduce the gains achieved with WOM-v codes. We discuss the tradeoffs between improved performance gains and reduced endurance gains using NR mode in detail in Section 4.2.3. Second, NR-mode require additional storage and memory by storing in a chunk's metadata the generation for each page. However, this added data is on the order of a few bits for each page, which seems negligible (2 bits per page) as compared to pre-existing in-memory metadata such as the logical to physical map (32-64 bits per page).

### 3.2.3 Adding QLC Support to FEMU

We use FEMU [17] to emulate the underlying SSD media. FEMU emulates the SSD in main-memory and adds predictable I/O latency to each I/O request to mimic a real Open Channel SSD device. In order to read or write a page, a specific number of reference voltages need to be applied to access the page. The number of reference voltages applied increases with the increase in flash density [14].

The main challenge in using FEMU for new generation SSD device emulation is that the existing FEMU emulator only supports an MLC SSD with 2 page levels. A page can either be an Upper or a Lower page with write latency of $850\mu s$ and $2300\mu s$ and the corresponding read latency of $48\mu s$ and $64\mu s$ respectively. FEMU also adds a constant NAND read, write and erase latency of $40\mu s$, $200\mu s$ and 2ms to each read, write and erase I/O request respectively.

We modify the default MLC configuration of FEMU's page layout in each chunk. Instead of having alternating upper and lower pages with varying latency in each chunk, we extend FEMU for QLC emulation: Each chunk in a QLC device has alternating Lower(L), Center-Lower(CL), Centre-Upper(CU), Upper(U) pages. A write latency of $850\mu s$, $2300\mu s$, $3750\mu s$ and $5200\mu s$, and read latency of $48\mu s$, $64\mu s$, $80\mu s$ and $96\mu s$ is applied to L,CL,CU and U pages respectively based on the number of reference voltages [14] required to read a specific page type.

### 3.2.4 Testbed for Future SSDs and Coding Schemes

Our WOM-v simulator is generic and can be used as a testbed for denser SSD or higher order coding schemes. In order to add a new WOM-v(k,N) coding scheme, first, the user has to provide a simple lookup table mapping each data word to a voltage level similar to WOM-v tables shown in Figure 2. Second, the user optionally sets an *ECC_threshold* value. Finally, the user specifies the latency of additional page levels for next generation SSDs in FEMU.

Our emulator can also be used standalone without any coding scheme. We have open-sourced our generic N-LC Simulator for more advanced coding and next generation SSD research [7].

## 4 Evaluation

In order to evaluate the gains of WOM-v(k,N) codes on QLC drives, it is important to evaluate the endurance gains and performance tradeoffs associated with WOM-v(k,N) codes. In particular, WOM-v(k,N) codes can potentially reduce the number of EUs by enabling overwrites between erases. However, WOM-v(k,N) also introduces space amplification during writes, which increase EUs and affect performance. Further, both read and write workflows in WOM-v(k,N) codes introduce read amplification for the device.

The goal of this section is to use our implementation to evaluate the EC reduction and the impact on performance for both micro-benchmarks and real world workload traces.

Since WOM-v(k,N) requires more space than NO_WOM configuration, there are two ways to compare them. The first option is to keep the size of the user facing logical address space constant, and increase the physical space allocated to WOM-v(k,N) configuration by a factor of N/k. For a fair comparison of EC reduction per hardware chip, we have to divide the resultant number of ECs encountered by NO_WOM code by N/k to account for additional space provided to the WOM-v(k,N) scheme.

A second more practical approach to evaluate SSD hardware is to keep the physical capacity of the hardware constant, and reduce the logical block address range by a factor of N/k for WOM-v(k,N) coding scheme. For example, we use the same 4 GB physical space in NO_WOM and WOM-v(k,N) configurations. For NO_WOM, the entire 4 GB logical address space is exposed to the application, while for all modes of WOM-v(2,4) codes (i.e, WOM-v(2,4), WOM-v(2,4)-GC-OPT and WOM-v(2,4)-NR) the range of logical address space on the same 4 GB drive is reduced to a 2GB logical space, and similarly for all modes of WOM-v(1,4) codes, the same drive is exposed as a 1 GB logical space. We assume the workload running on such a drive to have a logical space of 1 GB.

Under the same workload this setup leads to more empty space for NO_WOM configurations which acts as extra over provisioned space to reduce garbage collection overheads. However, we show that for the same physical device WOM-v(k,N) is able to successfully reduce the number of ECs in all our experiments with minor performance overheads.

In the rest of this section, we compare the default SSD configuration (NO_WOM) with the baseline implementation

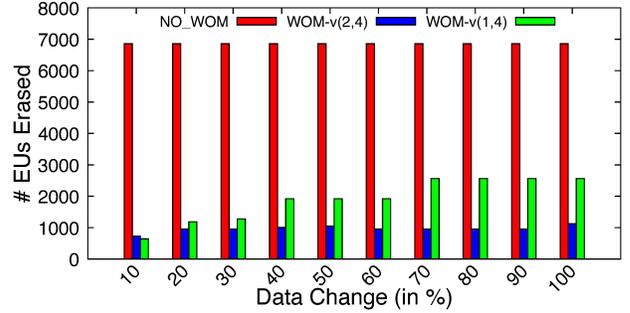

(a) *Effect of rate of data change on reduction in the number of EUs erased.*

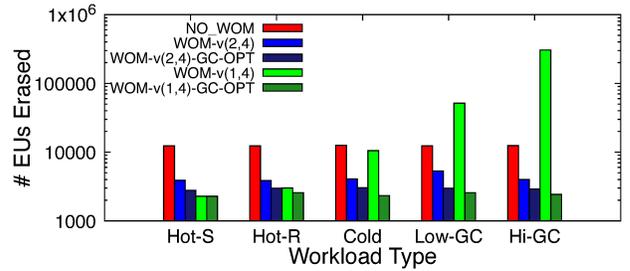

(b) *Effect of access pattern on reduction in the number of EUs erased.*

Figure 4: *Endurance measurement for microbenchmarks* of WOM-v(k,N) codes, garbage collection optimized (WOM-v(k,N)-GC_OPT) and performance optimized no-reads mode (WOM-v(k,N)-NR) implementations. We did not implement binary WOM(k,N) code in our emulator as the overall gains will be even lower than those obtained in No-WOM configuration (recall Section 2).

### 4.1 Micro Benchmarks

We use micro-benchmarks to study the impact of specific access patterns and data-block contents, on WOM-v(k,N) codes. Each micro-benchmark writes 25 GB data to 1 GB physical emulated drive containing 4 Parallel Units and 160 EUs.

#### 4.1.1 Effect of change in data buffer contents

The gain of the baseline WOM-v(k,N) codes depend on the data contents of the block that is overwritten to the existing data, due to the possibility of same generation transitions in WOM-v codes.

In this micro-benchmark, we fill the drive sequentially so no garbage collection is invoked. Once the drive is full, we flip a fraction of all bits in the data buffer, and fill the entire drive again with this modified data. We repeat this multiple times (Figure 4a X-Axis) and compare the number of EUs erased as the rate of change in data buffer contents increases.

Figure 4a shows the total number of EUs erased for varying data buffer contents during writes. We note that for all types of data buffer contents, WOM-v codes reduce the number of EUs erased compared to the NO_WOM configuration. However, the rate at which a cell reaches the maximum voltage level will be slower when there is smaller rate of data change between subsequent writes. For workloads with a higher amount of data change, the maximum voltage level will be reached faster, and so the EU erase gains will be lower. Unlike WOM-v codes, the NO_WOM configurations' EUs erased remain constantly at the higher end irrespective of the data buffer contents. The number of EUs erased in the NO_WOM configuration also remain constant as we run the same sequential pattern of write workload with different data contents.

#### 4.1.2 Effect of access Pattern

In order to measure the impact of access patterns, we create micro-benchmarks that invalidate previously written pages in a specific order to cause varying degrees of device write amplification: *Hot-S* keeps updating the same data sequentially, *Hot-R* updates the same data in a random order, *Cold* only updates a fraction of pages, *Low-GC* and *High-GC* generate medium and high amount of Garbage collection, respectively.

Figure 4b shows the number of EUs erased for different workload patterns. Across all benchmarks, we observe that WOM-v(1,2) codes significantly reduce the number of EUs erased. However, with higher number of overwrites due to GC, we see diminishing gains for WOM-v(1,4) code. Finally, GC_OPT mode is able to alleviate the problems incurred and maintain low write amplification overhead.

*Hot-S* keeps all pages "hot" i.e. uniformly accessed over the course of the benchmark. During garbage collection no pages from the previously written EU are garbage collected and hence we do not have any write amplification. We observe that *Hot-R* also does not create any garbage collection. Hence the pattern of write between two workloads does not impact the gains by WOM-v codes if the garbage collection thread writes minimal or no additional pages to the drive.

For the *Cold* benchmark, we observe a slight increase in the total number of EUs erased for the WOM-v(2,4) configuration and an order of magnitude increase in EU erases for the WOM-v(1,4) configuration as compared to *Hot-S* and *Hot-R* configurations. This is due to localized writes on only a subset of the device. We also observe that WOM-v(1,4)-GC-OPT and WOM-v(2,4)-GC-OPT continue to maintain lower write amplification for this benchmark as there is almost always a candidate EU with at-least a single programmable page, and hence it significantly reduces the number of valid pages from hot EUs that are relocated.

Low-GC creates a moderate amount of write amplification in the NO_WOM configuration. The WOM-v(2,4) configuration continues to outperform NO_WOM. But the WOM-v(1,4) configuration starts performing poorly as compared to the NO_WOM configuration. This is because additional writes generated space amplification in the WOM-v(1,4) configuration. Even with Low-GC, since there are continuous page invalidations, both WOM-v(2,4)-GC-OPT and WOM-v(1,4)-GC-OPT continue to reprogram an EU without relocating valid pages in a significant portions of reprogram operations, which keeps the write amplification relatively low.

| Source | # Traces | Medium | Year |
|---|---|---|---|
| Alibaba [18] | 814 | SSD | 2020 |
| RocksDB/YCSB Trace [25, 30] | 1 | SSD | 2020 |
| Microsoft Cambridge [22] | 11 | HDD | 2008 |
| Microsoft Production [15] | 9 | HDD | 2008 |
| FIU [16] | 7 | HDD | 2010 |

Table 1: *Historical HDD and recent SSD based block traces*

High-GC benchmarks cause severe write amplification due to high influx of valid pages recycled during garbage collection. This causes WOM-v(1,4) to perform two orders of magnitude worse than NO_WOM. However, even in High-GC mode, WOM-v(1,4) GC_OPT continues to find EUs that have intermediate pages available for reprogramming for a significant fraction of reprogram operations, and hence the write amplification remains consistent over the course of the workload run.

#### 4.1.3 Conclusions from micro benchmarks

We conclude that across all data write patterns, WOM-v codes are highly effective in reducing the number of EUs erased. For workloads where similar or incremental data is overwritten on the device, huge gains are possible.

WOM-v(2,4) codes are highly robust to different kinds of workload patterns. For workloads that exhibit increased garbage collection, WOM-v(k,N)-GC-OPT codes continue to maintain near constant EUs erased even for artificial, extremely-high garbage collection workload.

### 4.2 Real World Evaluation

#### 4.2.1 Setup

**Trace Selection:** Table 1 shows a summary of 844 real world traces from 5 different sources. We shortlist and present 10 write-based traces and 6-read-based traces representing each source. The write-based workloads have higher number of writes than reads and help us to measure the endurance improvement (Section 4.2.2) and write performance tradeoffs (Section 4.2.3). The read-based workloads help us better understand the impact of WOM-v codes on read performance (Section 4.2.3).

The selected traces have at-least 1 million and at most 50 million page writes. We also select traces with varying number of unique block accesses and varying number of updates per unique block. We choose enterprise workloads (except [30]) instead of synthetic benchmarks for our evaluation for two reasons. First, the enterprise workloads exhibit significantly different workload characteristics than the TPC and

FIO benchmarks, that are specifically designed to stress the system under test. Second, enterprise workloads show variation in usage over time, for example due to diurnal patterns.

There are a few limitations of using real-world workload traces. First, there are no publicly available block traces where TRIM information is available. We consider a block to only be invalidated if it is overwritten. Hence our estimate of gains using WOM-v codes are rather conservative. When considering the addition of pages invalidated with the TRIM command, there will be more opportunity for reprogramming physical pages that contain TRIM'ed logical pages in WOM-v codes. Moreover, we will have lower write amplification induced by garbage collection that will favour WOM-v codes.

Second, block-traces do not have any information about the buffer contents. We fill each block with random data. This does not impact NO_WOM configuration, but impacts the WOM-v configurations and higher gains could be possible if the data is more uniform causing less state change. The Microsoft Cambridge (MC-stg, MC-rsch), Florida International University (FIU-online,FIU-websearch) and Microsoft Production (MP-backend, MP-authentication) are older traces collected on HDDs. The Alibaba workloads (Alibaba-316, Alibaba-746, Alibaba-4) and RocksDB traces represent workloads that are aware of the underlying SSD device.

The traces require pre-processing for various reasons. The real world workloads are captured on different sized SSDs and HDDs. Further, traces only capture a subset of the logical block numbers of the entire drive. Also, our FEMU emulator models the underlying SSD storage in main memory. so the emulated drive size is much smaller than the original SSD or HDD on which the trace was captured.

To standardize these traces, we pre-process each real-world trace and reduce it to a format that can be run on a constant sized 16GB FEMU Based LightNVM emulator. During trace reduction, we ensure that the access pattern of each page in the original and reduced trace remains the same. We reduce the I/O size proportionately, if required and ensure the distribution of page update frequencies across the trace remains the same.

**FEMU+LightNVM Drive Setup:** When configuring the QLC drives for trace-based FEMU+LightNVM experiments, for each trace we set up a drive with the same *physical* capacity (same number of cells) for all NO_WOM and WOM-v(k,N) configurations. We size the capacity of the drive such that the NO_WOM drive would be half full (half of its capacity is used), which is common for real-world drives [8, 19, 20]. For each workload, we compute the number of unique blocks accessed, and set up a QLC drive that is twice the size of the total number of unique blocks accessed. We note that the logical address space for all WOM-v(2,4) setups is half the logical address space of NO_WOM. Similarly, the logical address space for WOM-v(1,4) setups is one-fourth the logical address space of NO_WOM.

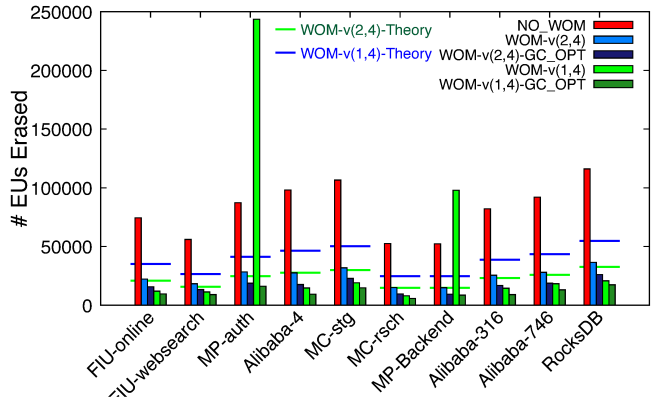

Figure 5: *Reduction in the number of Erase Units (EUs) erased using different configurations of WOM-v codes.*

### 4.2.2 Reduction in number of erase operations

This section uses our experimental testbed to evaluate the improvements in drive lifetime that different WOM-v codes achieve across a variety of workloads. In particular, we are interested in comparing the number of Erase Unit (EU) erases required to run the different workloads on a standard NO_WOM QLC drive and WOM-v enabled QLC drives, as these EU erases directly affect drive lifetime.

We first consider the question of whether a baseline WOM-v implementation, without any of the optimizations we propose in Section 3.2.2, can reduce the EU erases compared to a NO_WOM QLC drive. Toward this end, we observe in Figure 5, that the WOM-v(2,4) code reduces the total number of EU erased consistently by a factor of at least 3 (between 68% to 71% reduction) across all traces. Results for the WOM-v(1,4) code are a bit more mixed: In most cases WOM-v(1,4) further reduces the number of EU erasures, providing an additional factor of two reduction on top of WOM-v(2,4). Two notable exceptions are the MP-auth and MP-Backend traces, where the WOM-v(1,4) code erases more EUs than the NO_WOM configuration. This shows that there are certain workloads where the reduction in erase operations that WOM-v codes offer through overwrites between erases does not make up for the write amplification that comes with a higher rate code. (Recall that in a WOM-v(1,4) code each data bit gets encoded in 4 coded bits). The two MP workloads are more skewed in their popularity distribution, which results in higher garbage collection cost, and the added write amplification in WOM-v(1,4) codes gets amplified.

The poor performance of the WOM-v(1,4) codes for GC-intensive workloads motivates us to study the impact of the GC_OPT setting that we introduced in Section 3.2.2 to reduce write amplification during WOM-garbage collection. Figure 5 shows that enabling GC_OPT does indeed reduce the number of erases. With the GC_OPT setting both WOM-v(2,4)-

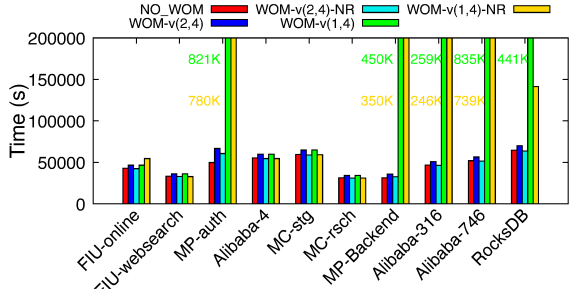

Figure 6: *No Read Mode Performance*

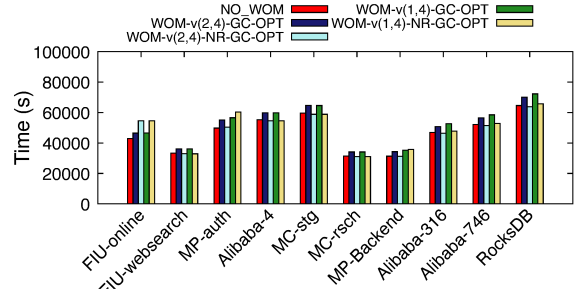

Figure 7: *No Read Mode with GC_OPT performance*

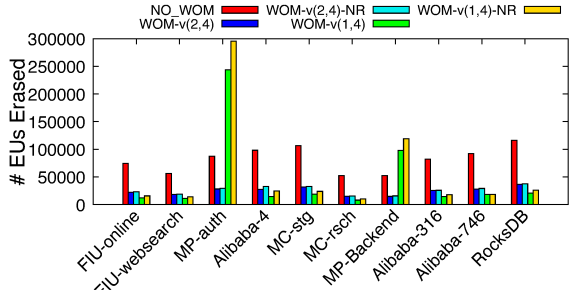

Figure 8: *No Read Mode endurance*

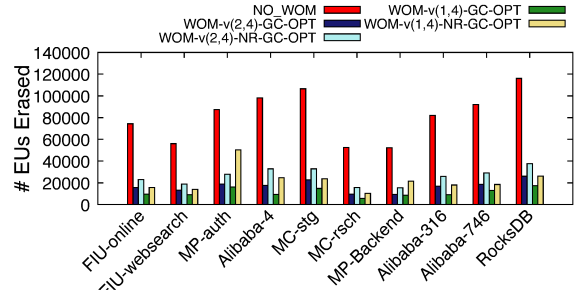

Figure 9: *No Read Mode with GC_OPT Mode endurance*

*Figure 6 and 7 show performance improvement using NR Mode for WOM-v(k,N) baseline and WOM-v(k,N) GC_OPT modes, respectively, and Figure 8 and 9 compare endurance gains.*

GC_OPT and WOM-v(1,4)-GC_OPT improve significantly over NO_WOM across *all workloads*, including the two MP workloads where the baseline WOM-v(1,4) did not improve over NO_WOM. The number of erases is reduced by 77-83% for WOM-v(2,4)-GC-OPT and 82-91% for WOM-v(1,4)-GC-OPT compared to NO_WOM. This translates to $4.4 - 11.1\times$ reduction in erase operations for real-world workloads.

It is interesting to compare the results from our experimental evaluation to the simplistic back-of-the envelope results in [12], which estimates reductions in the number of erase operations by $5\times$ for WOM(1,4) and of $3.5\times$ for WOM(2,4). While we observe results in this range for some workloads, we are able to draw more differentiated conclusions. Improvements vary by workload and in particular GC-intensive workloads see significant improvements only if care is taken to optimize the GC process.

We also compared our results with theoretical estimates based on an analytical model presented by Yaakobi et al. in [29]. The authors in [29] derived an analytical model to analyze the gain that a WOM coding scheme can achieve in terms of reducing the expected number of erase operations, while considering the trade-off between the write amplification induced by the coding rate, including its effect on the garbage collection, and the number of reprogram operations the WOM code can perform between two erase operations. We use the number of write cycles and the coding rate in our different WOM-v code settings and the fullness of the drive based on our traces to determine the erase factor derived in [29] equation (3) and then convert the erase factor to the number of Erase Units erased. The corresponding lines are labelled WOM-v(k,N)-Theory in Figure 5. We observe that the experimental results of WOM-v codes achieve considerably larger reductions in erases than what was predicted by the analytical model. The reason is most likely that the analysis in [29] relies on several simplifying assumptions in order to be tractable, including for example an assumption that page invalidations happen uniformly at random. These observations underline the importance of trace-driven experimental evaluation of the benefits of WOM codes in SSD drives.

In conclusion, we demonstrate that WOM-v codes can greatly reduce the number of erase operations across a wide range of workloads. For higher rate codes, such as WOM-v(1,4) codes, and workloads that result in high garbage collection it is important to employ GC_OPT mode to limit write amplification during garbage collection.

#### 4.2.3 Performance Evaluation

This section evaluates the impact of WOM-v(k,N) coding on a drive's performance. Section 3.2.2 already discussed how the read-before-write requirement of standard WOM codes can impact write performance and presented the WOM-v_NR optimization to eliminate this requirement (potentially at the cost of reduced endurance gains). A second potential source of performance impact is the generally higher write workload generated by WOM-v(k,N) as each write adds additional (N/k) data to be written. The goal of this section is to quantify the performance impacts of WOM-v(k,N) coding across different workloads.

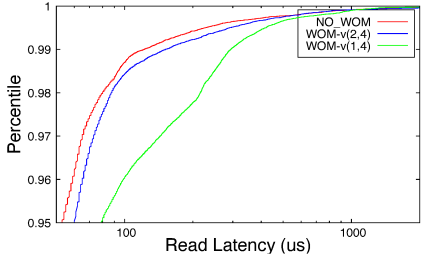
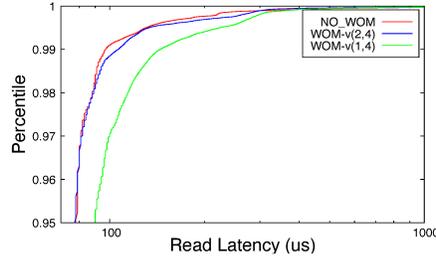
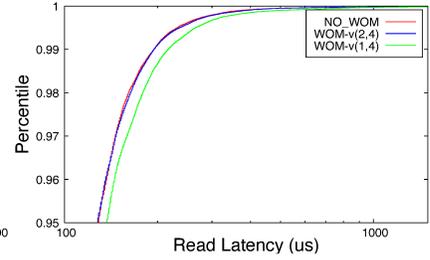

Figure 10: *MSR-Cambridge Web1*  Figure 11: *MSR-Production DAPL*  Figure 12: *Alibaba 3*

*Figure 10, 11 and 12 show read tail latency of MSR-Cambridge Web1, MSR-Production Display Ads Payload and Alibaba-Server 3 workload with read:write ratios of 11:9, 4:3 and 5:6 respectively.*

**Average performance / throughput** Figure 6 compares the cumulative time to run our ten workloads on a NO_WOM drive versus WOM-v(k,N) drives with and without NR mode enabled. A larger cumulative time means lower system throughput and higher average request latency.

We observe that for the WOM-v(2,4) configuration the performance overheads are quite small, in the 3-8% range, compared to the NO_WOM configuration. Adding the NR mode optimization eliminates those overheads and leads to performance comparable to a NO_WOM configuration. At the same time we observe in Figure 8 that enabling NR mode for the WOM-v(2,4) configuration leads to only minimal reduction in the endurance gains compared to plain WOM-v(2,4).

We also evaluate whether adding the GC_OPT optimization, either in isolation, or in combination with NR mode will reduce performance impacts, in particular for the WOM-v(1,4) scheme, which has high overheads in the baseline implementation. The results are shown in Figure 7.

We find that enabling both GC_OPT and NR mode greatly reduces performance overheads for the WOM-v(1,4) scheme, bringing performance within 0-8% of the NO_WOM configuration. At the same time we observe in Figure 9 that when enabling GC_OPT and NR mode we achieve large endurance gains over a NO_WOM drive.

**Read Tail Latency** Finally, while the above discussion focused on average performance, we also considered tail latency, which is particularly important for read requests. Figure 10, 11 and 12 show the tail latency of read requests across three traces that we chose because they were read intensive. For all three cases, we observe that the 95'th percentile tail latency is 0.6-7% for NO_WOM and WOM-v(k,N) baseline encoding schemes, and that there is no large tail latency introduced due to the use of WOM-v(k,N) code.

### 4.3 Comparison with MLC Drives

Given that increases in drive endurance from WOM codes come with space overheads, an interesting question is how the WOM-v endurance/space tradeoff compares to simply using MLC technology. MLC cells have higher PE cycle limits than QLC drives, but are less space efficient. In particular, if we reduce a QLC drive to an MLC drive with the same number of physical cells, we would lose 50% of logical capacity (2 bits per cell instead of 4), which is identical to the logical capacity loss when applying WOM-v(2,4) codes to a QLC drive. In exchange for the 50% capacity loss, the MLC drive's endurance will increase because the PE cycle limit of a cell increases from 3K for a QLC drive to 10K for an MLC drive [1–4], and the WOM-QLC drive's endurance will increase compared to QLC due to overwrites between erases.

The goal of this section is to compare the endurance of an MLC drive to that of a WOM-v(2,4) QLC drive with the same logical capacity and the same number of physical cells. Endurance is the amount of user data that can be written before the PE cycle limit is reached. We can use our results from Section 4.2.2 to estimate endurance for a WOM-v(2,4) QLC drive for a given workload, based on the number of erases observed for the experiment with the corresponding trace and the amount of user data written in the trace. We also ran experiments for all workloads on an MLC drive with the same physical capacity and recorded those numbers.

Figure 13 compares the ratio of the endurance of an WOM-QLC drive and the endurance of an MLC drive for different workloads based on the methodology described above. We observe that for the same write pattern, the endurance of the WOM-QLC drive exceeds the endurance of the MLC drive in all scenarios (ratio larger than 1). This is the case even for the workloads with higher garbage collection where improvements from WOM codes were lower than for other workloads. On average the improvement in endurance is a factor of 3.5x for GC_OPT mode and 2.4x for GC_OPT-NR Mode, with negligible performance overheads as compared to a NO-WOM QLC drive using the NR Optimization. In summary, we conclude that WOM-v codes can significantly improve drive endurance compared to standard MLC as well as QLC drives.

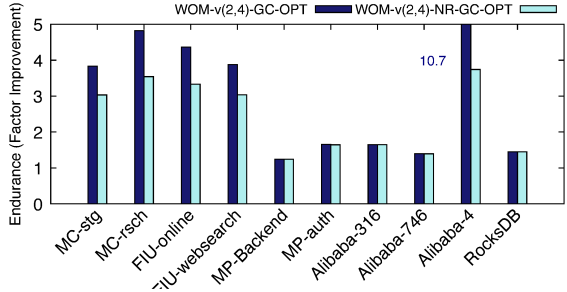

Figure 13: *The ratio of write endurance of WOM-v based QLC drives in GC_OPT and NR-GC_OPT mode and the write endurance of an MLC drive with the same logical capacity and the same number of physical cells. For all workloads, WOM-v(2,4) enabled QLC drives provide better endurance than MLC drives.*

## 5 Related Work

The closest to this work is our own earlier workshop paper [12]. In [12], we introduced the idea of WOM-v codes and used a preliminary back of the envelope analysis to demonstrate their potential in theory. The goal of this paper is to study whether and how we can bring the advantages of WOM-v codes into practice where real world factors such as workload access patterns, device realities (write amplification, parallelism, etc) as well as its performance implications are taken into account. We provide a full implementation of WOM-v codes based on FEMU and LightNVM, including a discussion of implementation challenges and some optimizations we made, and an evaluation using micro-benchmarks as well as a trace-driven evaluation. We demonstrate that higher order WOM-v(k,N) baseline codes may lead to drastically higher endurance and performance overheads for real world, high garbage collection workloads. To alleviate this issue, we propose GC_OPT, a novel garbage collection mechanism that continues to provide endurance and performance gains for WOM-v(k,N) codes.

Other prior work [27, 28, 31] studies the use of traditional bit-based WOM codes for MLC and TLC flash. However, those studies focus on binary WOM coding scheme that cannot scale on higher density drives such as QLC drives. Further, in Section 2, we note that binary WOM codes do not provide gains for QLC drives. Our focus is a voltage-based coding scheme designed for high-density SSD drives.

Margaglia et. al. [21] present hardware limitations in MLC drives that reduce the gains achievable by WOM codes in practice. In particular, they observe that certain bit transitions are not feasible within the flash hardware. Such restrictions could appear for QLC drives as well. However, WOM-v(k,N) codes would avoid these restrictions as they operate directly based on voltage levels. Furthermore, WOM-v(k,N) codes present a family of codes - if a few state transition restrictions were imposed by the underlying hardware, one can use a lower code ratio WOM-v(k,N) code that is more suitable to the flash hardware restrictions. In [13], the authors show how erase operations are detrimental to NAND-flash. They recommend differential levels of programming and erase to normalize the amount of charge issued to erase a NAND flash chip. [13] is complementary to our work and both can be used together to improve overall flash lifetime.

## 6 Summary and Implications

Below we summarize our findings and their implications.
- WOM-v codes reduce the number of erase operations by 4.4-11.1x for QLC SSDs. This directly improves the lifetime of flash media and reduces the overhead of purchasing, replacing and restoring data from an older drive.
- Even for workloads that exhibit high-write amplification, WOM-v codes present a novel opportunity to delay garbage collection in GC_OPT mode until the entire erase unit is no longer programmable. Such optimization is possible *only* in WOM-v codes where data can be overwritten without erasing previous data on a SSD.
- Even for performance-critical workflows deployed on high-density SSDs, drive endurance can be enhanced to a significant degree using WOM-v codes without compromising on performance using the No-Reads optimization.
- The space amplification of WOM-v codes can be a cause of concern for SSDs that exhibit high space utilization. For such drives, we recommend routinely transitioning between WOM-v codes with different code rates based on the drive fullness or the space requirements of the workload. We leave evaluation of such dynamically transitioning coding schemes based on space-utilization as part of future work.
- While this paper demonstrates the effective use of WOM-v codes and discusses the implementation, evaluation, and optimization of WOM-v codes in the context of QLC SSDs, WOM-v codes can be easily extended to higher density, future generation of SSDs such as PLC SSDs and more sophisticated coding schemes.

## Acknowledgements

We thank our USENIX FAST '22 reviewers and our shepherd Rob Ross for their detailed feedback and valuable suggestions. We thank Charles Xu for his help with trace selection and processing. This work was supported by an NSERC Discovery Grant and an NSERC Canada Research Chair.